\documentclass[preprint2]{aastex}

\usepackage[english]{babel}
\usepackage{amsmath}
\usepackage{amssymb}
\usepackage{url}
\usepackage[retainorgcmds]{IEEEtrantools}

\newcommand{\Ks}[1]{K_{\mathrm{s} #1}}

\shorttitle{Double HB in NGC\,6440 and NGC\,6569}
\shortauthors{Mauro et~al.}

\begin{document}

\title{Double Horizontal Branches in NGC\,6440 and NGC\,6569 unveiled by the VVV Survey\thanks{Based on observations gathered with ESO-VISTA telescope (proposal ID 172.B-2002).}}

\author{Francesco Mauro\altaffilmark{a}, Christian Moni Bidin\altaffilmark{a}, Roger Cohen\altaffilmark{a}, Doug Geisler\altaffilmark{a}, Dante Minniti\altaffilmark{b,c}, Marcio Catelan\altaffilmark{b,c}, Andr\'e-Nicolas Chen\'e\altaffilmark{a,d}, Sandro Villanova\altaffilmark{a} }
 
\altaffiltext{a}{Departamento de Astronom\'ia, Universidad de Concepci\'on, Casilla 160-C, Concepci\'on, Chile, email: fmauro@astroudec.cl}
\altaffiltext{b}{Departamento de Astronom\'ia y Astrof\'isica, Pontificia Universidad Cat\'olica de Chile, Casilla 306, Santiago, Chile}
\altaffiltext{c}{The Milky Way Millennium Nucleus, Av. Vicu\~{n}a Mackenna 4860, 782-0436 Macul, Santiago, Chile}
\altaffiltext{d}{Departamento de de Fis\'ica y Astronom\'ia, Universidad de Valpara\'iso, Av. Gran Breta\~na 1111, Playa Ancha, Casilla 5030, Chile}

\begin{abstract}
We report the discovery of a peculiar horizontal branch (HB) in NGC\,6440 and NGC\,6569, two massive and metal-rich Galactic globular clusters (GGCs) located in the Galactic bulge, within 4\,kpc from the Galactic Center.
In both clusters, two distinct clumps are detected at the level of the cluster HB, separated by only $\sim 0.1$ magnitudes in the $\Ks{}$ band.
They were detected with IR photometric data collected with the ``VISTA Variables in the V\'ia L\'actea'' (VVV) Survey, and confirmed in independent IR catalogs available in the literature, and HST optical photometry.
Our analysis demonstrates that these clumps are real cluster features, not a product of field contamination or interstellar reddening.
The observed split HBs could be a signature of two stellar sub-populations with different chemical composition and/or age, as recently found in Terzan\,5, but it cannot be excluded that they are caused by evolutionary effects, in particular for NGC\,6440.
This interpretation, however, requires an anomalously high helium content ($Y>0.30$).
Our discovery suggests that such a peculiar HB morphology could be a common feature of massive, metal-rich bulge GGCs.
\end{abstract}

\maketitle

\section{Introduction}
\label{s:intro}

Our understanding of the complexity of Galactic Globular Clusters (GGCs) has impressively expanded in the last decade, propelled by the discovery that they can host multiple populations of stars with a different chemical enrichment history \citep{Piotto2005}.
The classical text-book definition of GGCs as prototypes of a simple stellar population, i.e. a chemically homogeneous aggregate of coeval stars, is now out-dated.
While a certain degree of inhomogeneity of light chemical elements is observed in nearly all GGCs \citep{Carretta2009}, a spread in iron content is a characteristic restricted to only a few very massive objects \citep{Freeman1975,Yong2008,Cohen2010}.
\citet{Ferraro2009} discovered two horizontal branches (HBs) in the Bulge GGC Terzan 5, separated by 0.3~magnitudes in the $\Ks{}$ band.
The existence of multi-modality in the morphology of HBs has been known for nearly four decades \citep{Harris75}, and has been associated with the presence of multiple stellar populations since shortly thereafter \citep{Rood85}.
However, to date, Terzan 5 is the only GC known to have two distinct HBs.
The two features in Terzan\,5 have a different spatial distribution, the brighter one being more centrally concentrated, more metal rich \citep[ $\Delta${[Fe/H]}$\sim 0.5$~dex,][]{Origlia2011}, and possibly helium enhanced \citep{DAntona2010} and/or younger \citep{Ferraro2009}.
\citet{Lanzoni2010} confirmed that Terzan\,5 is more massive than previously thought, and it could be the relic of a Bulge building block.

In this Letter, we show evidence that the Bulge GGCs NGC\,6440 and NGC\;6569 host split HBs, similar to that of Terzan\,5.
NGC\,6440 is a high-metallicity \citep[{[Fe/H]}$\approx -0.5$,][]{Origlia2008} cluster, located 8.5~kpc from the Sun and only 1.3~kpc from the Galactic center \citep[][2010 edition, H10]{Harris1996}.
NGC\,6569 is slightly less metal-rich \citep[{[Fe/H]}$\approx -0.79$,][]{Valenti2011} and is found at a distance of 10.9~kpc from the Sun and 3.1~kpc from the Galactic center (H10).
Both NGC\,6440 and NGC\,6569 are among the ten most luminous of the 64 GGCs located within 4\,kpc from the Galactic center.

\section{Observations and reductions}
\label{s:data}

The ``VISTA Variables in the V\'ia L\'actea'' (VVV) Survey  \citep{Minniti2010} is one of the six ESO Public Surveys operating on the 4-meter Visible and Infrared Survey Telescope for Astronomy (VISTA).
VVV is scanning the Galactic bulge and the adjacent part of the southern disk
($-65\leq l\leq-10$, $-2\leq b\leq +2$),
in five near-IR bands ($YZJHK_\mathrm{s}$) with the VIRCAM camera \citep{Emerson2010}, an array of sixteen 2048$\times$2048~pixel detectors with a pixel scale of $0\farcs 341/pix$.
VVV images extend four magnitudes deeper and exhibit increased spatial resolution \citep{Saito2010} versus Two Micron All Sky Survey \citep[2MASS,][]{2MASS}, which is particularly important for mitigating contaminated photometry in crowded regions near the Galactic center.

\begin{figure}[ht!]\begin{center}
\epsscale{1.}
\plotone{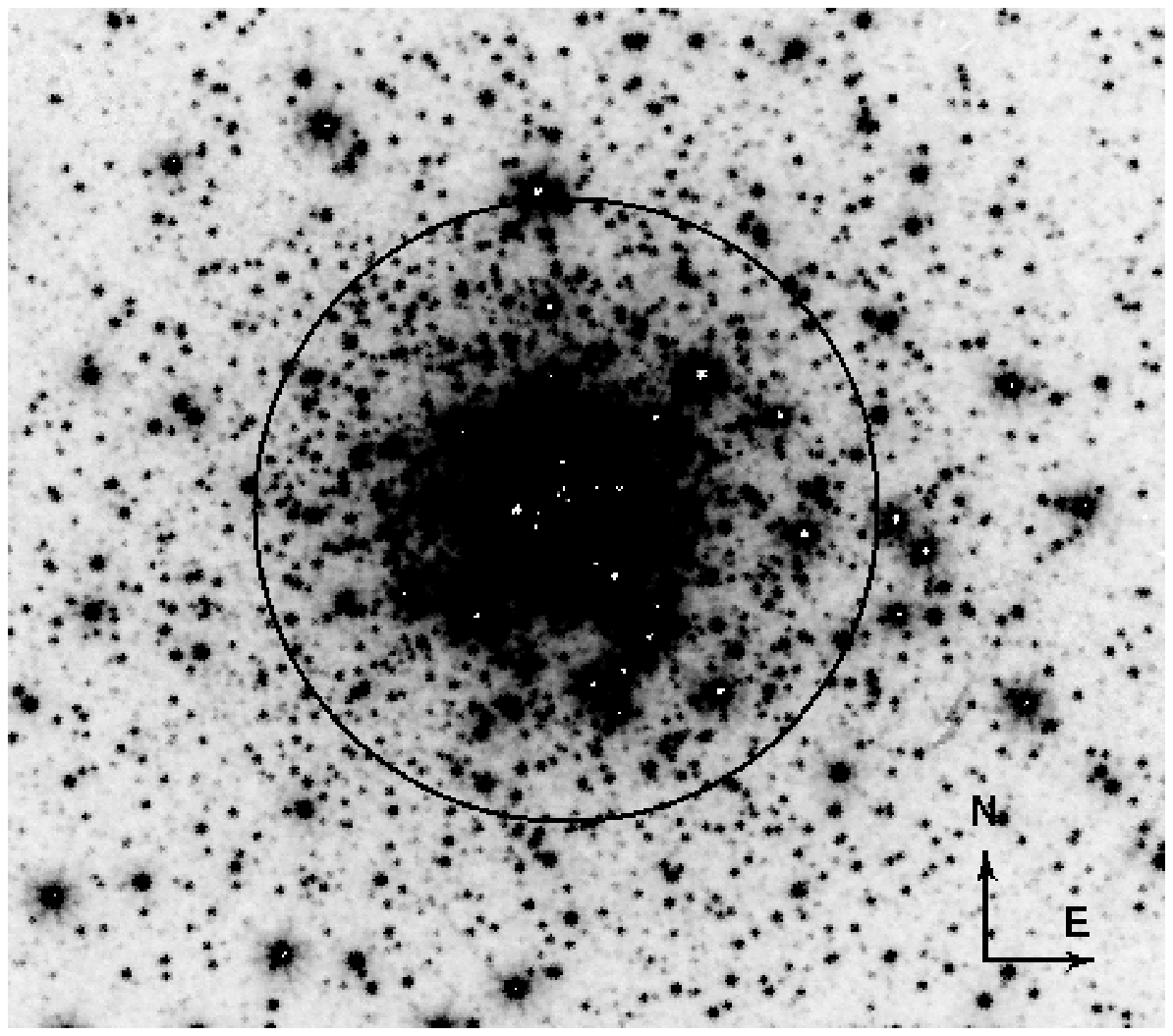}
\plotone{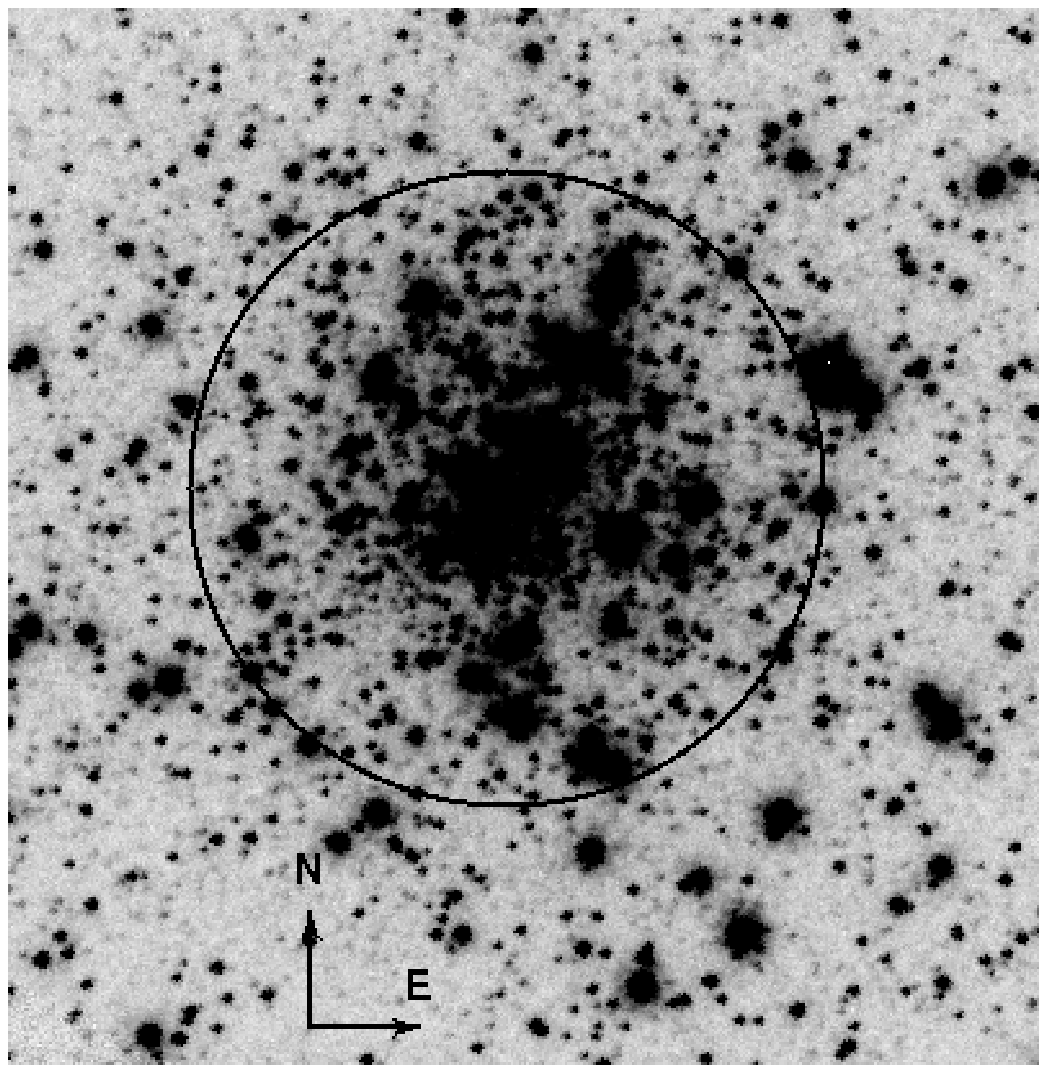}
\caption{$2\arcmin\times 2\arcmin$ linear gray-scale $\Ks{}$-band VVV images of NGC\,6440 (upper panel) and NGC\,6569 (lower panel). The circle indicates the $0\farcm 7$ inner limit of photometric completeness commented in the text.}
\label{fig:field}
\end{center}\end{figure}

We retrieved from the Vista Science Archive website\footnote{\url{http://horus.roe.ac.uk/vsa/}} the VVV images of the two GGCs, pre-reduced at the Cambridge Astronomical Survey Unit (CASU)\footnote{\url{http://casu.ast.cam.ac.uk/}} with the VIRCAM pipeline \citep{Irwin04}.
The selected data consist of four frames, sampling twice each point in an area of $17\arcmin\times 22\arcmin$ around the GGCs, in each of the $ZYJH\Ks{}$ filters, plus 17 and 11 additional epochs in the $\Ks{}$ passband \citep{Saito2012} for NGC\,6440 and NGC\,6569, respectively.
The VVV images of the two clusters, extracted from a single $\Ks{}$ frame, are shown in Figure \ref{fig:field}.

The PSF-fitting photometry was obtained with the VVV-SkZ\_pipeline (VSp, Mauro et al. {\it submitted}), code based on DAOPHOT and ALLFRAME \citep{DAOPHOT,ALLFRAME} procedures, optimized for the VVV data.
The photometry was tied to 2MASS JHKs standards, as described in \citet{Moni2011} and \citet{Chene2012}.
Combining all the 36 and 24 $\Ks{}$ measurements, the final photometric errors were 0.003 and 0.008 mag respectively at the brightness level of the cluster HB.
The $\Ks{}$ errors for NGC\,6569 are costant with distance, while in NGC\,6440 they increase up to $0.005$ mag for $r<0\farcm 7$.
The completeness of our photometry is heavily affected by crowding in the inner $0\farcm 7$ of both clusters, as can be appreciated in Figure \ref{fig:field} and \ref{fig:Rdistr}.
Due to incompleteness, $\sim80\%$ of the detected HB stars stay outside this problematic inner region, where crowding is not a significant issue.

\section{Results}
\label{s:res}

\begin{figure}[ht!]\begin{center}
\epsscale{1.}
\plotone{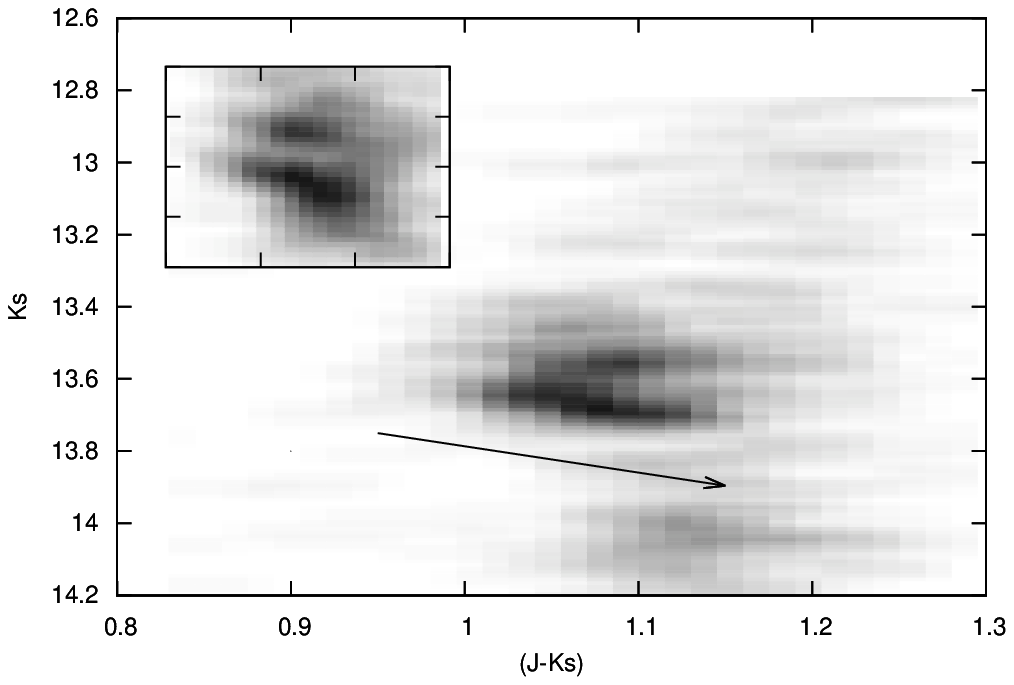}
\plotone{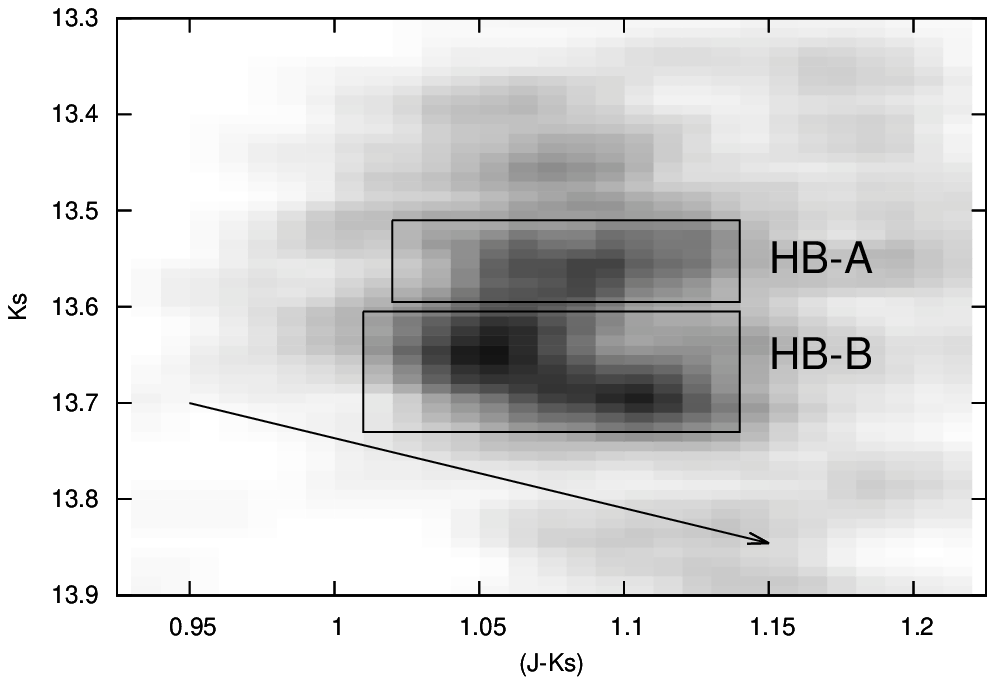}
\caption{$(J-\Ks{};\Ks{})$ Hess diagrams of HB area of NGC\,6440.  The dereddened version is presented in the insert plot in upper panel.  
In lower panel, the groups HB-A and HB-B discussed in the text are spotlighted. 
The arrows show the effect of the reddening.}
\label{fig:6440hessdiagr}
\end{center}\end{figure}

\begin{figure}[ht!]\begin{center}
\epsscale{1.}
\plotone{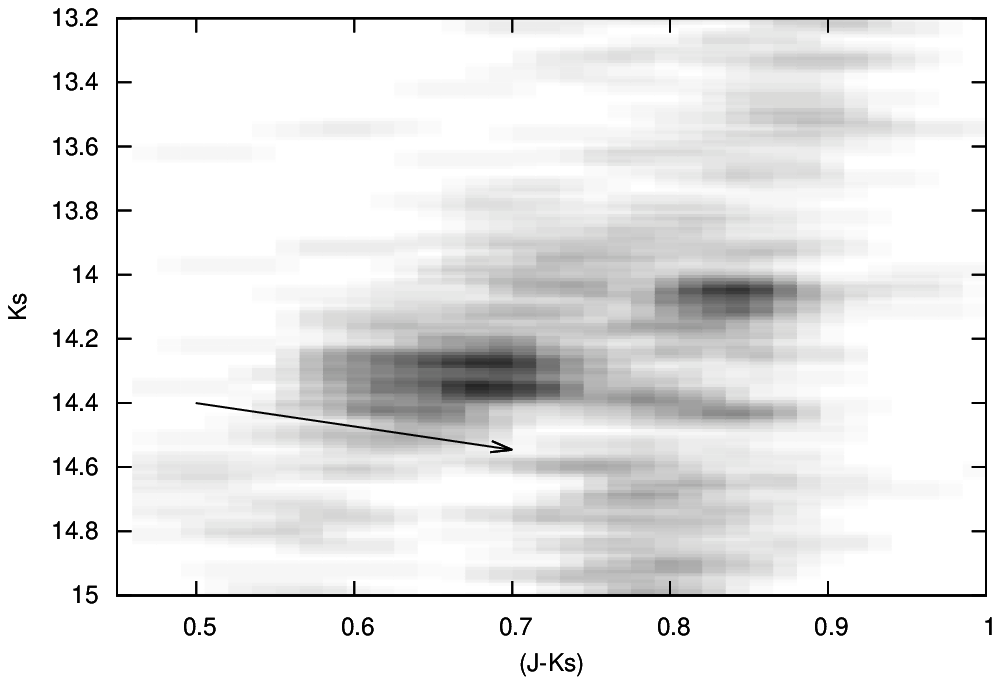}
\plotone{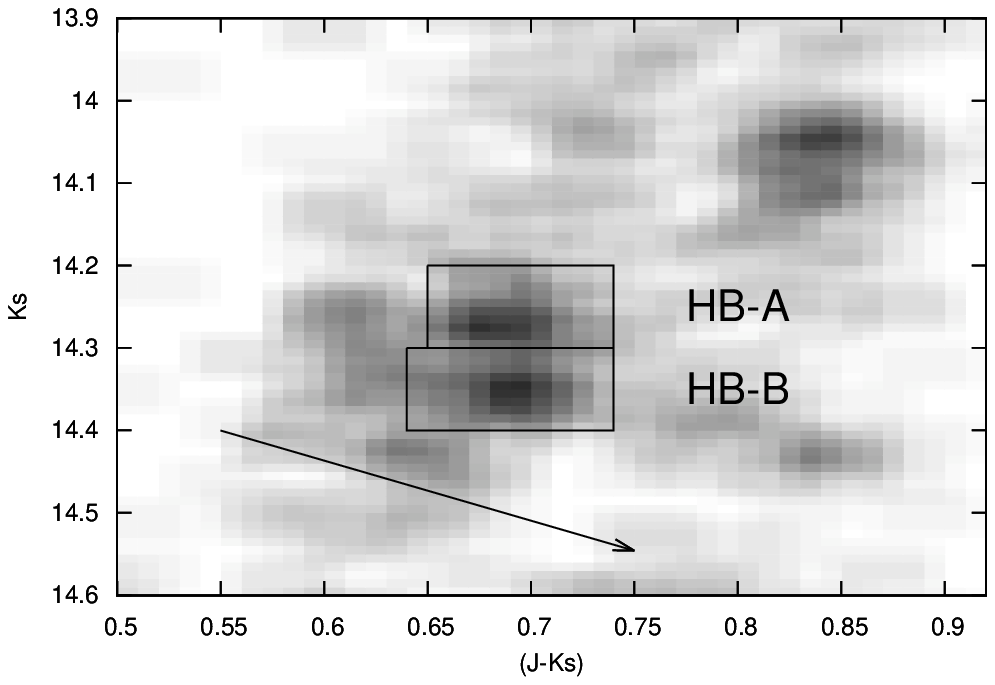}
\caption{Same as Figure \ref{fig:6440hessdiagr} for NGC\,6569.}
\label{fig:6569hessdiagr}
\end{center}\end{figure}

The $(J-\Ks{};\Ks{})$ Hess diagrams (HD) for the stars detected within $1\farcm 83$ and $1\farcm 66$ from the cluster center, respectively, are shown in Figure~\ref{fig:6440hessdiagr} and \ref{fig:6569hessdiagr}.
They were obtained calculating the number of stars in a bin of width $0.06\;mag$ in $J-\Ks{}$ ($0.04\;mag$ in the lower panels) and $0.04\;mag$ in $\Ks{}$, moved along the axis with steps of $0.015\;mag$ in color ($0.01\;mag$ in the lower diagram) and $0.01\;mag$ in magnitude.
The HDs of both GGCs reveal a peculiar HB morphology, with two distinct clumps separated by $\sim 0.1\;mag$.
They will be referred to as HB-A and HB-B (lower panels of Figures \ref{fig:6440hessdiagr} and \ref{fig:6569hessdiagr}) for the brighter and fainter one, respectively.
The overdensity observed at redder color is the red giant branch (RGB) bump: in fact, \citet[][V05]{Valenti2005} found $K_{s,bump}$=14.08 in NGC\,6569, and their $K_{s,bump}$-[Fe/H] relation predicts $K_{s,bump}\approx$14.1 in NGC\,6440, in good agreement with our data.

\defcitealias{Valenti2005}{V05}

To verify if both HBs are real and belong to their host cluster, we checked the data for stochastic fluctuations as a cause of the overdensity, and analyzed their spatial distribution as a function of the central distance.
Furthermore, we compared our data with the IR photometry of \citet[V04]{Valenti2004} and  \citetalias{Valenti2005}, and with the optical HST data from \citet{Piotto2002}.
\defcitealias{Valenti2004}{V04}

\paragraph{Dereddening.}
We used the maps from \citet{Gonzalez2011} to correct for reddening.
They reveal that $E(J-\Ks{})\sim0.5-0.7$ in the $r=1\farcm 8$ field of NGC6440 under analysis.
The case of NGC6569 is much less extreme, with $E(J-\Ks{})=0.20-0.24$.
The HD of NGC\,6440 shows a clear improvement (see Figure \ref{fig:6440hessdiagr}), with the two features less blurred and HB-B still presenting a slope.
For NGC\,6569 the dereddened HD is approximately similar to the raw one, as expected.

\paragraph{Checking for stochastic variation.}
We reran the procedure on four subsets of the original data, each one containing the $ZY$ and $JH\Ks{}$ data, but different $\Ks{}$ epochs: one subset included only the first epoch, while a unique set of three epochs were used in each of the three following subsets.
The declared photometric errors in $\Ks{}$ passband vary from $0.007-0.009\;mag$ to $0.003-0.005\;mag$ at the level of the HB.
Comparing the $(J-\Ks{};\Ks{})$ HDs, obtained with the previous spacial selection and sampling procedure, both GGCs always exhibit a split HB, with only negligible differences in their morphology.
As an additional test, we checked the HBs of other GGCs in the VVV, namely NGC\,6380, NGC\,6441, NGC\,6528 and NGC\,6553, finding no evidence of a split or peculiar HB.

\begin{figure}[ht!]\begin{center}
\epsscale{1.}
\plotone{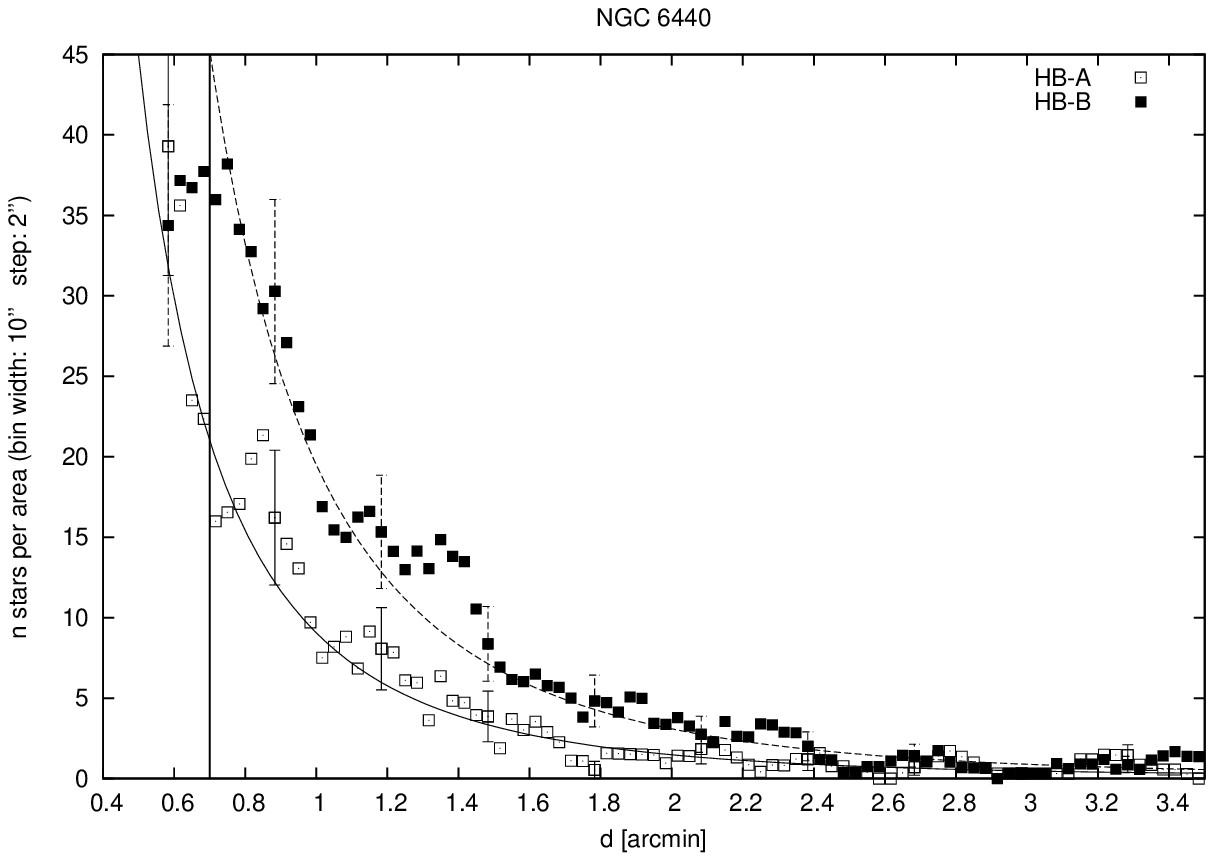}
\plotone{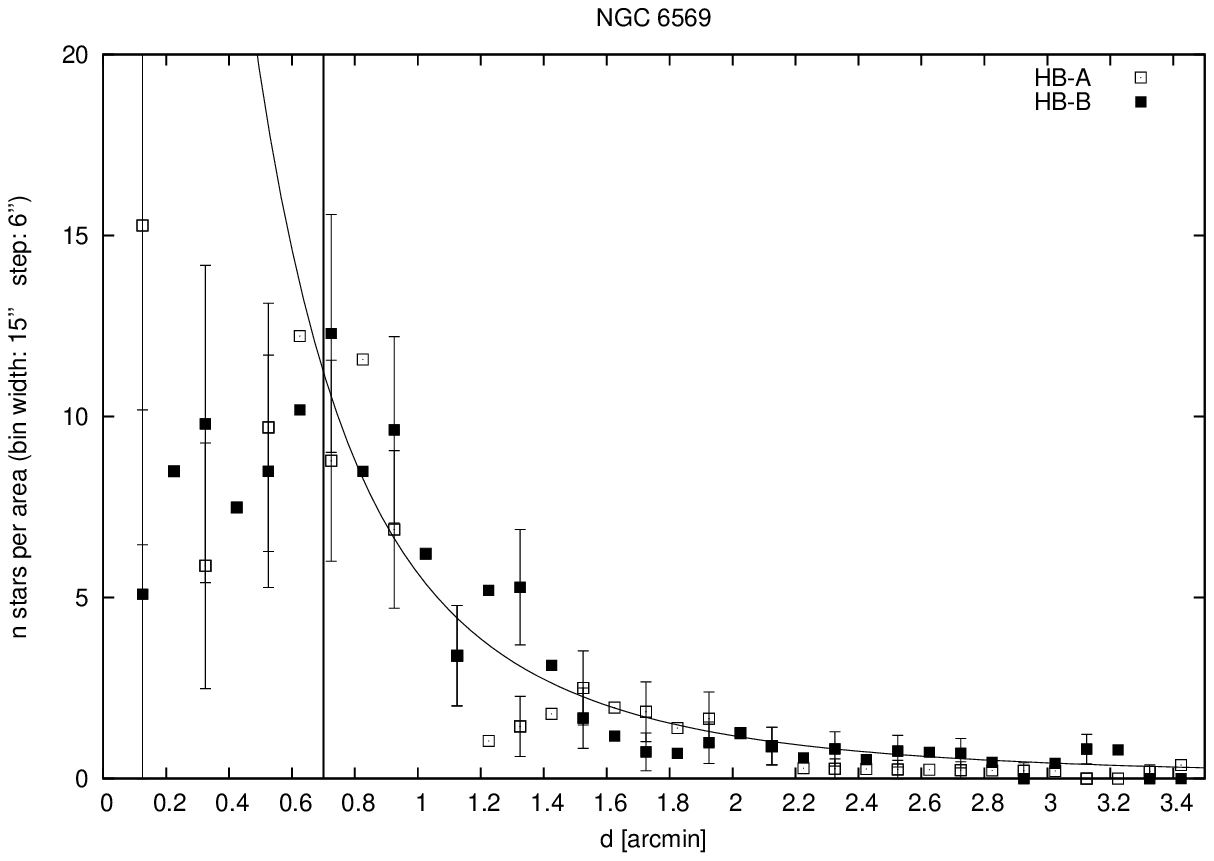}
\caption{The distributions of the stellar density as function of the distance (bin width of 10\arcsec moved with steps of 2\arcsec) for the HB-A (solid line) and HB-B (dashed line) features.
The error bars are plotted only every $0\farcm 3$ and every $0\farcm 25$ for NGC\,6440 and NGC\,6569, respectively; the King profile fits are overplotted.
The vertical lines indicate the limit for the incompleteness.
}
\label{fig:Rdistr}
\end{center}\end{figure}

\paragraph{Field Contamination and Spatial Distribution.}
We checked the field contribution to the HDs, selecting an annular region with the same area of the previous selection, but just outside the tidal radius.
The HDs of the field are barely populated at the HB location, and the field contamination is negligible.
This results is evident even in Figure \ref{fig:Rdistr}, where the number counts drop to near-zero levels at large distances from the cluster centers.

The behavior of the stellar densities (SDs) with distance  $r$ from the center is shown in Figure \ref{fig:Rdistr} for the two features highlighted in the lower panels of Figures \ref{fig:6440hessdiagr} and \ref{fig:6569hessdiagr}.
The SDs (stars per $arcmin^2$) were calculated with a bin width of $10\arcsec$ moved at steps of $2\arcsec$ for NGC\,6440, while for NGC\,6569 we used the values of $15\arcsec$ and $6\arcsec$, respectively.
The SDs of the two groups steeply decay at increasing radii, in both GGCs, and their members are distributed on the CCD with circular symmetry.
The radial profile of the two features in NGC\,6569 is identical.
The HB-B group in NGC\,6440 is more populated than the brighter HB-A by a factor of two, but a Kolmogorov-Smirnov test reveals that their radial behavior coincides also in this case.
The stellar counts in the inner $0\farcm 7\simeq 5 r_c$ of NGC\,6440 (where $r_c$ is the core radius from H10) are incomplete because of crowding.
The photometry of NGC\,6569 is also incomplete for $r<0\farcm 7\simeq 2 r_c$. 

The radial profile of NGC\,6569 was fit with a  \citet{King1962} profile of the form
\begin{equation}\label{eq:king1}
 f(r)=k\left\lbrace\left[1+\left(\frac{r}{r_c} \right)^2  \right]^{-\frac{1}{2}}\!\!\!\!-\left[1+\left(\frac{r_t}{r_c} \right)^2  \right]^{-\frac{1}{2}} \right\rbrace^2+F,
\end{equation}
where $k$ is a scale parameter, $r_t$ is the tidal radius and $F$ the field contribution.
For NGC\,6440, we used the approximation for $r\gg r_c$
\begin{equation}\label{eq:king2}
 f(r)=kr_c^2\left(\frac{1}{r}-\frac{1}{r_t}\right)^2+F \,.
\end{equation}
For NGC\,6440, the fit
leads to $r_{t,A}=5\farcm 1\pm0\farcm 7$ and $r_{t,B}=5\farcm 2\pm 0\farcm 5$, consistent with $r_t=5\farcm 84$ quoted by H10.
The core radius and the scale parameter cannot be separated and estimated individually.
The SD of the two features in NGC\,6569 are compatible with $r_c=0\farcm 35$ and $r_t=7.15$ (H10).

\paragraph{Comparison with previous photometry.}
We matched our VVV photometry of  NGC\,6440 and NGC\,6569 with the catalogs of \citetalias{Valenti2004} and \citetalias{Valenti2005}, respectively.
The photometry of \citetalias{Valenti2004} is based on observations with the near-IR camera IRAC2@ESO/MPI 2.2m,
covering a $250\arcsec\times 250\arcsec$ field centered on the cluster.
Similarly, the photometry of \citetalias{Valenti2005} was performed on data collected with the near-IR camera SOFI@ESO/NTT, 
, covering a $300\arcsec\times 300\arcsec$ field centered on the cluster.
The estimated internal photometric errors are lower than 0.03 mag.
Both photometries were calibrated onto the 2MASS photometric system and astrometrically corrected by using the 2MASS catalog.

\begin{figure}[ht!]\begin{center}
\epsscale{1.}
\plotone{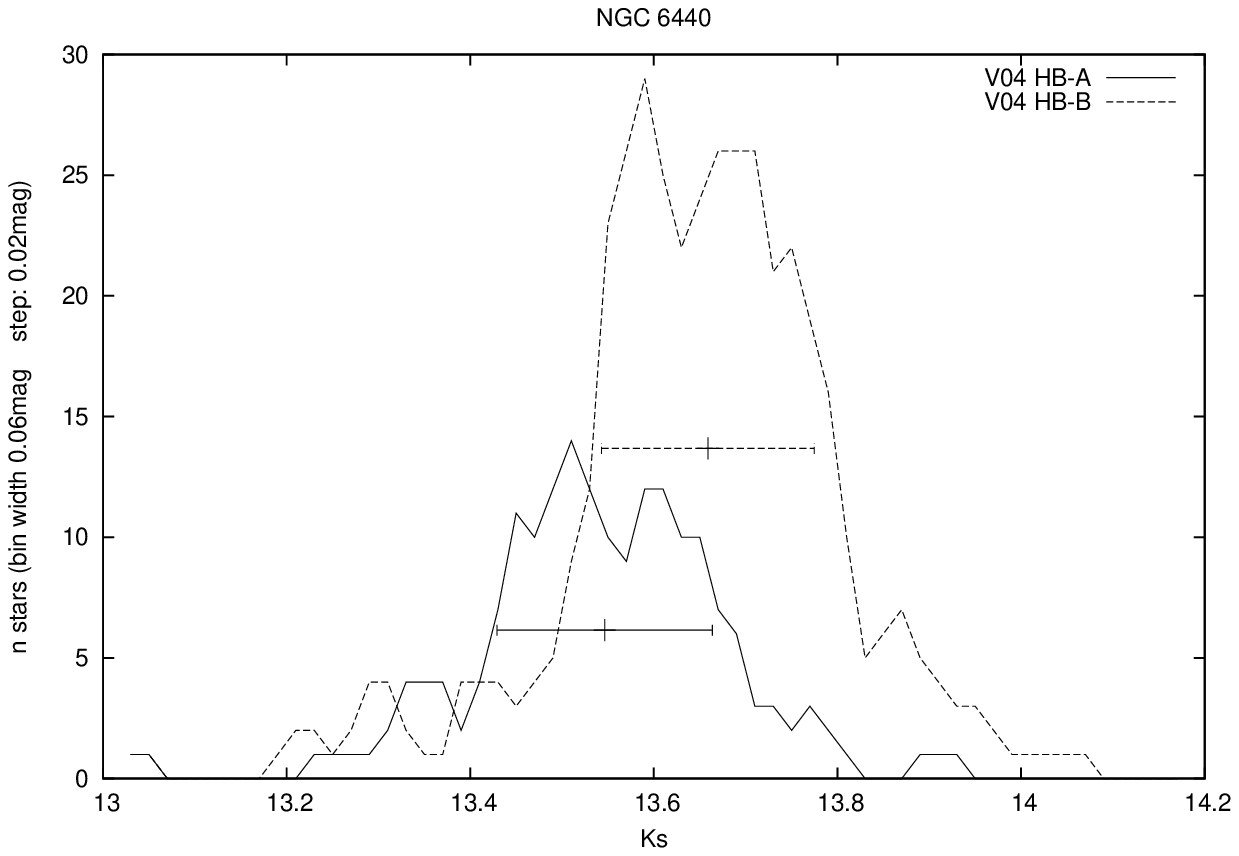}
\plotone{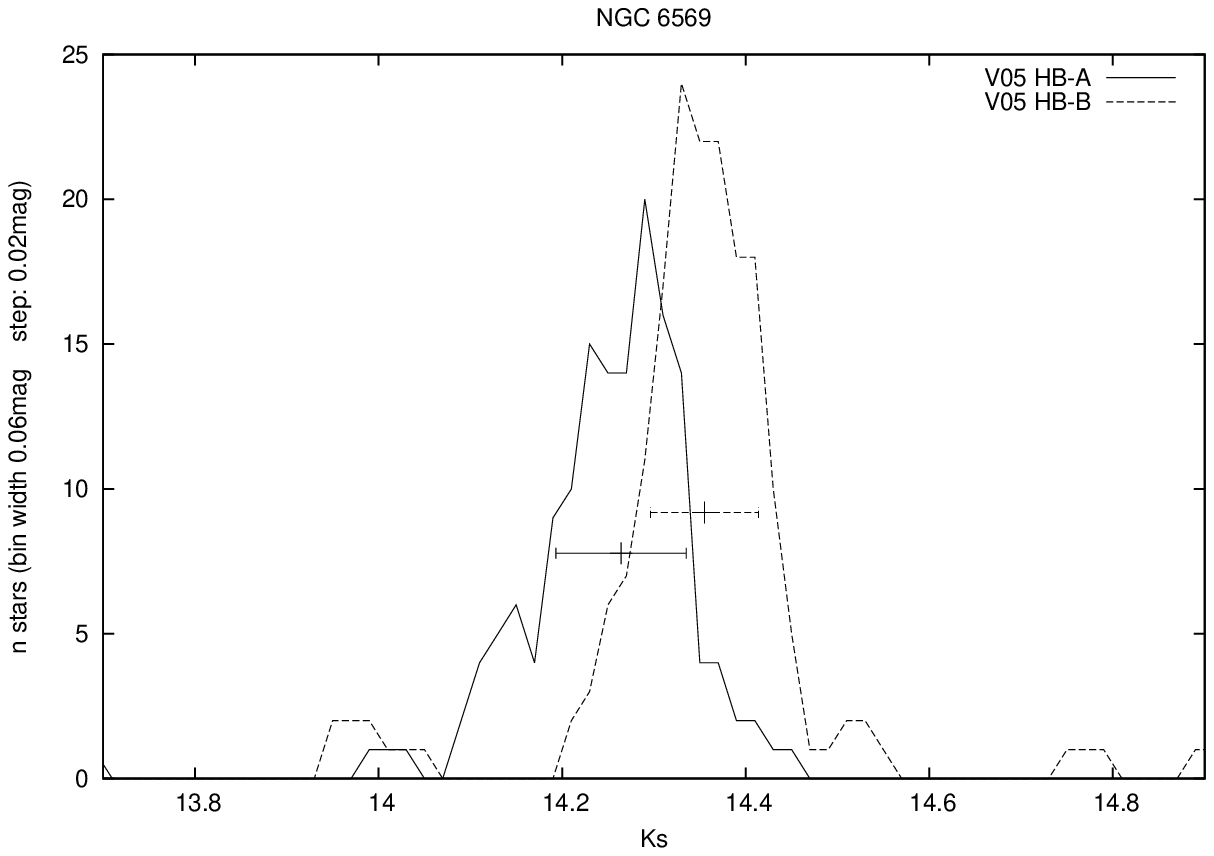}
\plotone{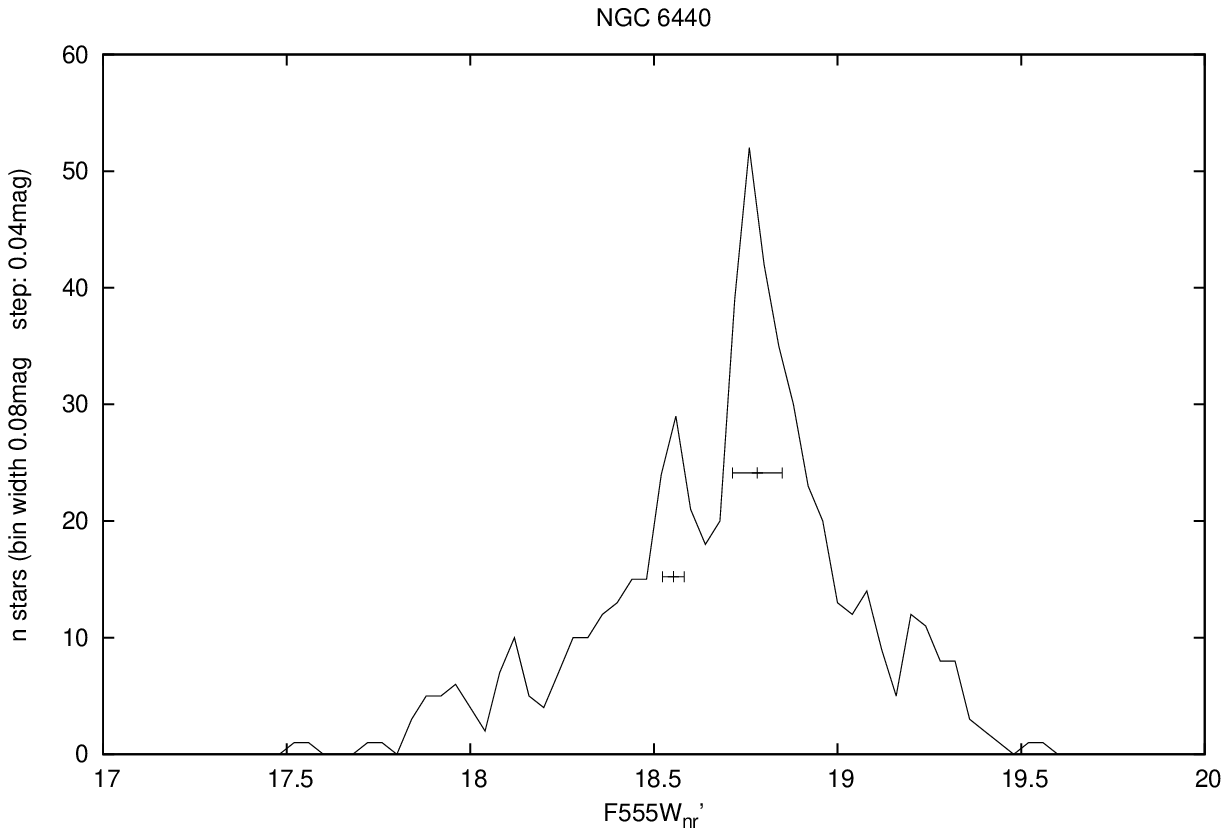}
\caption{Luminosity distribution in $\Ks{V04}$ and $\Ks{V05}$ passband (in upper and middle frame for NGC\,6440 and NGC\;6569, respectively) for the matched stars belonging to the two features HB-A and HB-B of the respective GGCs.
Luminosity distribution of the values $F555W_{nr}'$ for NGC\,6440 in the lower frame.
The center and the sigma resulting from the Gaussian fits are also shown.}
\label{fig:Val}
\end{center}\end{figure}

For both GGCs, the luminosity distributions in the $\Ks{}$ magnitudes of the \citetalias{Valenti2004} and \citetalias{Valenti2005} catalogs for the matched stars do not show a clear bimodal distribution.
However, when the stars belonging to the HB-A and HB-B groups are identified, their luminosity distributions are different, as shown in Figure \ref{fig:Val}.
For NGC\,6440, the Gaussian fits of the two distributions are centered at $\Ks{,V04}=13.55$ and $\Ks{,V04}=13.66$ for HB-A and HB-B, respectively, with a dispersion of $\sigma=0.12$, while in our VVV photometry the values are $\Ks{VVV}$=13.55 and 13.67, respectively, with a dispersion of $\sigma=0.03$.
Analogously, we find $\Ks{,V05}=14.26$ and $\Ks{,V05}=14.36$, respectively, for the HB-A and HB-B clumps in NGC\,6569, with a dispersion of $\sigma=0.07$, and $\Ks{VVV}$=14.26 and 14.35 for the same features in our photometry, with a dispersion of $\sigma=0.02$.
Thus, the mean magnitude of the clumps of both GGCs is identical in VVV and Valenti et al.'s catalogs, but the separation is four times more statistically significant in the VVV data.
This result proves that the HBs of both GGCs are intrinsically split in magnitude, with a brighter and  a fainter part that remain separated even in \citetalias{Valenti2004} and \citetalias{Valenti2005} photometry, respectively, once the stars are identified.

%
%

The strong differential reddening affecting the region of NGC\,6440 causes the HB to be strongly tilted at optical wavelengths, hence a simple luminosity distribution does not show a bimodal behavior.
For this reason, we analyzed the HST optical data of \citet{Piotto2002} projecting the position of each HB star along the HB slope, according to the equation

\begin{IEEEeqnarray*}{rl}
\label{eq:P8}
F555W_{nr}'= & F555W_{nr}-a[(F439W_{nr}-F555W_{nr})\\ 
 &  -(F439W_{nr}-F555W_{nr})_0], \IEEEyesnumber\\
\end{IEEEeqnarray*}
 
where $a=3.7$ is the slope of the HB and $(F439W_{nr}-F555W_{nr})_0=2.2$ is the HB mean color.
As advised by the authors, the magnitudes adopted were those not corrected for reddening (\emph{nr}).
To avoid contamination from RGB stars, we selected only the sources with $18.2\le F555W_{nr}\le 19.1$ and $2\le (F439W_{nr}-F555W_{nr})\le 2.4$.
The distribution of $F555W_{nr}'$, calculated with bin width of $0.08$ mag and step of $0.04$ mag (see Figure \ref{fig:Val}), reveals a clear double peak separated by $\sim 0.23$ mag, with the fainter peak 1.6-1.7 times more populated.
A similar procedure was performed for NGC\,6569 also, 
but we were not able to disentangle any bimodality.

\section{Discussions and Conclusions}

The analysis of VVV data reveals that the HB of the GGCs NGC\,6440 and NGC\,6569 is split into two distinct clumps.
This behavior is not introduced by stochastic fluctuations of the density in the CMD, or induced by photometric errors, as it is found to be identical in four independent subsets of data.
Field contamination is not the cause either, because the members of both the HBs are distributed with spherical symmetry with respect to the cluster center, and their density steeply decays with distance.
The separation in NGC\,6440 is even cleaner after applying a differential reddening correction, while in NGC\,6569 it remains similar, presenting lower differential reddening.
This HB split is, however, not found in the same VVV data of four less massive Bulge GGCs.

The magnitude difference between the two HB clumps is only $\sim 0.08-0.1$ magnitudes in $\Ks{}$, smaller than in Terzan\,5 by a factor of three.
It is thus not surprising that this HB split passed unnoticed in previous investigations, also considering the strong differential reddening affecting the NGC\,6440 field \citep[$\Delta E(J-\Ks{})=0.2$,][]{Gonzalez2011}.
We find that the dichotomy is blurred by observational errors in the IR photometry of \citetalias{Valenti2004} and \citetalias{Valenti2005}, but the two features are clearly separated even in their data, once their members are identified in their catalog.
The HST optical data of NGC\,6440 from \citet{Piotto2002} show two peaks separated by $0.23$ mag in the corrected magnitude $F555W_{nr}'$ defined in Equation~\ref{eq:P8}.

The fainter HB of NGC\,6440 is bluer than the brighter one.
This resembles what was previously found in Terzan\,5 \citep{Ferraro2009,Lanzoni2010}.
In addition, the fainter HB of NGC\,6440 is more populated than the brighter one by about a factor of two, slightly higher than what was found in the central regions of Terzan\,5 \citep[$\sim 1.6$,][]{Ferraro2009}.
However, these results are not directly comparable, because our photometry is incomplete in the inner $0\farcm 7$ of NGC\,6440.
HST data suggest that in the central region this ratio could be lower ($\sim 1.6$), as expected if, analogously to Terzan 5, the brighter HB was more centrally concentrated.
On the other hand, we did not detect any difference in the radial density profile of the two clumps, so the issue remains open.

The two HBs of Terzan\,5 are associated with two sub-populations of different metallicity, with the brighter HB being richer in iron by $\sim 0.5$ dex \citep{Ferraro2009,Origlia2011}.
According to \citet{Salaris2002}, a difference of $\sim$0.3~dex would be expected, if the observed HB splits are interpreted only in terms of metallicity.
Nevertheless, this is only a rough upper limit, because differences in helium content and age also  can contribute to cause the same split.
\citet{Origlia2008} measured the metallicity of ten stars in NGC\,6440, finding a dispersion of only 0.06~dex, compatible with observational errors.
However, their targets are mainly located within $0\farcm 7$ of the center, where the HB-A members could be few if the population ratio is constant at all radial distances.
Hence, Origlia et al.'s sample likely contains only a small quantity of HB-A stars, and their results are insufficient to exclude a metallicity spread in this cluster.

Contrary to the case of NGC\,6440 and Terzan\,5, the two groups identified in the HB of NGC\,6569 have approximately the same color and the same radial distribution, the fainter HB B being 1.3 times more populated than the other clump.
\citet{Valenti2011} measured the metallicity of six stars in this cluster, finding a bimodal distribution with two groups separated by $\sim 0.08$ dex.

It is possible that the peculiar HB morphology discovered in NGC\,6440 is a pure evolutionary effect, and not a signature of the presence of sub-populations.
In fact, wedge-shaped HBs are predicted under special circumstances, as depicted in Fig.~4 of \citet{Catelan1996} and Fig.~1 of \citet{Dorman1989}, and statistical effects could lead to the actual bimodal distribution of HB magnitudes.
These features are found in the luminosity-temperature plane, but the simulated optical CMDs reveal only a clump at the bluer (and brighter) end of the sloped HB, at variance with what is observed in HST data, while the behavior of these features in the IR bands has not been simulated.
Hence, this interpretation seems unlikely, but it cannot be excluded and represents an intriguing possibility.
In fact, the high metallicity alone cannot explain the formation of a wedge-like HB, and a very high helium abundance ($Y>0.30$) is required.
Such a He-enriched field population has been recently suggested in the bulge \citep[e.g.,][]{NatafGould2012}, although at higher metallicities.
Hence, interpreting the HB morphology of NGC\,6440 as an evolutionary effect implies that the helium content of this cluster must be anomalously high, actually higher than what is predicted at [Fe/H]=$-0.5$ by the models of Bulge chemical enrichment \citep[e.g.,][]{Catelan1996}.
The split HB observed in NGC\,6569, whose components are well separated and with a narrow color spread, is very different to the simulated HBs of \citet{Catelan1996} and \citet{Dorman1989}.
The interpretation of these features as an evolutionary effect induced by a high helium content is unlikely.

The fainter HB of NGC\,6440 (HB-B) is tilted, the brightness of its stars increasing at bluer colors.
This is clearly visible in Figure~\ref{fig:6440hessdiagr}, where we indicate the direction of the reddening in the IR bands from \citet{Catelan2011} for comparison.
Dereddening the photometry, the slope is still present, but the map resolution of $1\arcmin$ does not permit strong claims.
The tilt is more pronounced in the optical HST photometry of \citet{Piotto2002}, where we measured a slope $\Delta (V)/\Delta (B-V)\approx$3.7, which is higher than the standard reddening law $R_V= A_V/E(B-V)\approx$3.1.
Moreover, the optical extinction is non-standard toward the Galactic bulge, and $R_V$ can be as low as $\sim$2.5 \citep{Nataf2012}.
In conclusion, the slope of the HB in NGC\,6440 is directed approximately aligned with the reddening vector, but it is steeper than the expectations of interstellar reddening.
This behavior was already observed in NGC\,6388 and NGC\,6441 \citep{Sweigart1998,Busso2007}, two other massive, metal-rich Bulge GGCs, and it was attributed to an anomalously high helium content \citep{Caloi2007}.
Very interestingly, the HB stars of NGC\,6388 could show the same peculiar properties observed in $\omega$\,Centauri \citep{Moehler2006,MoniBidin2011}, the most famous cluster hosting a complex mix of sub-populations with different chemical enrichment histories.

Our results indicate that Terzan~5 is not a unique object.
A complex HB morphology could be a relatively common feature among metal-rich, massive Bulge GGCs.
This is not detected in less massive objects (e.g. NGC\,6528, NGC\,6553), nor in equally massive but metal-poor Bulge GGCs, such as M\,22 and M\,28, whose HB is very extended toward the blue.
The large metallicity spread observed in Terzan\,5 \citep{Origlia2011} is also present in M22 \citep{Marino2009}.
Further investigations are needed to unveil if the HB splits reflect the presence of two stellar populations with different chemical composition and/or age.

\acknowledgments
We gratefully acknowledge support from the Chilean {\sl Centro de Astrof\'\i sica} FONDAP No.15010003 and the Chilean Centro de Excelencia en Astrof\'\i sica y Tecnolog\'\i as Afines (CATA) BASAL PFB-06/2007 .
ANC also gratefully acknowledges support from Comite Mixto ESO-Gobierno de Chile.
This project is supported by the Chilean Ministry for the Economy, Development, and Tourism's Programa Iniciativa Cient\'{i}fica Milenio through grant P07-021-F, awarded to The Milky Way Millennium Nucleus; by Proyecto Fondecyt Regular \#1110326; and by Proyecto Anillo ACT-86.

\end{document}